\documentclass[namedreferences]{solarphysics}

\usepackage[hyperref,optionalrh,showbiblabels]{spr-sola-addons} 
\usepackage{epsfig}          
\usepackage{graphicx}        
\usepackage{amssymb}        
\usepackage{color}           
\usepackage{breakurl}        
\usepackage{multirow}
\usepackage[urlcolor=blue,breaklinks]{hyperref}

\hypersetup{
	colorlinks = true,
	linkcolor = blue,
	anchorcolor = blue,
	citecolor = blue,
	filecolor = blue,
	urlcolor = blue
}

\ifx \doiurl    \undefined \def \doiurl#1{\href{http://dx.doi.org/#1}{\textsf{DOI}}}\fi
\ifx \adsurl    \undefined \def \adsurl#1{\href{http://adsabs.harvard.edu/abs/#1}{\textsf{ADS}}}\fi
\ifx \arxivurl  \undefined \def \arxivurl#1{\href{http://arxiv.org/abs/#1}{\textsf{arXiv}}}\fi

\chardef\us=`\_

\begin{document}
	
	\begin{article}
		\begin{opening}
			
			\title{Forward Modeling of the Type III Radio Burst Exciter}
			
			\author[addressref={aff1},email={pjzhang@mail.ustc.edu.cn}]{\inits{P.J.}\fnm{Peijin }~\lnm{Zhang}\orcid{orcid.org/0000-0001-6855-5799}}
			\author[addressref={aff1,aff2},corref,email={cbwang@ustc.edu.cn}]{\inits{C.B.} \fnm{Chuanbing}~\lnm{Wang}\orcid{orcid.org/0000-0001-6252-5580}}
			\author[addressref={aff1}]{\inits{L.}\fnm{Lin}~\lnm{Ye}\orcid{orcid.org/0000-0001-8611-5818}}
			\author[addressref={aff1,aff2}]{\inits{Y.M}\fnm{Yuming}~\lnm{Wang}\orcid{orcid.org/0000-0002-8887-3919}}
			
			
			\address[id=aff1]{CAS Key Laboratory of Geospace Environment, School of Earth and Space Sciences, University of Science and Technology of China, Hefei, Anhui 230026, China}
			\address[id=aff2]{CAS Center for the Excellence in Comparative planetology, Hefei, Anhui 230026, China}
			
			\runningauthor{P.J. Zhang et al.}
			\runningtitle{Forward Modeling Type III burst}
			
			\begin{abstract}
				In this work, we propose a forward modeling method to study the trajectory and speed of the interplanetary (IP) Type-III radio burst exciter. The model assumes that the source of an IP Type-III radio burst moves outward from the Sun following the Parker spiral field line. Using the arrival time of the radio waves at multiple spacecraft, we are able to determine the trajectory of the radio source in the Ecliptic plane, and its outward speed, as well as the injection time and longitude of the associated electron beam near the solar surface that triggers the Type-III radio burst. For the application of this method, we design a system to gather the arrival time of the radio wave from the radio dynamic spectra observed by \textit{Solar Terrestrial Relations Observator} (STEREO)/WAVES and \textit{Wind}/WAVES. Then, the system forward models the trajectory and speed of the radio burst exciter iteratively according to an evaluation function. Finally, we present a survey of four Type-III radio bursts that are well discussed in the literature. The modeled trajectories of the radio source are consistent with the previous radio-triangulation results, the longitude of the associated active region, {or the location of Langmuir waves excited by the electron beam.}
				
			\end{abstract}
		\end{opening}
		
		\section{Introduction}
		\label{S-Introduction} 
		
		A Type-III solar radio burst is excited by the energetic electrons injected into the open magnetic-field lines rooted near or in solar active regions. This was demonstrated by observations of the \textit{Orbiting Geophysical Observatory} (OGO)-5 showing that Type-III radio bursts are often associated with electron beams \citep{alvarez1972evidence}. More recently, the observational statistics from the \textit{3D Plasma Analyzer} (3DP) instrument on board \textit{Wind} shows that 98.75\,\% of solar electron events are associated with Type-III radio bursts \citep{wang2012statistical}. {These non-thermal electrons can generate the radio waves \textit{via} some coherent emission processes such as the plasma emission mechanism \citep{ginzburg1958possible,reid2014review} or the cyclotron maser mechanism \citep{wu2002generation,wang2015scenario,chen2017self}.} 
		
		The study of Type-III bursts can help us understand the acceleration process of energetic electrons during solar activity and electron transport in the solar atmosphere and interplanetary space \citep{reid2018solar,chen2013tracing,chen2018magnetic}. According to the classic plasma-emission mechanism, the non-thermal electron beams first excite Langmuir waves, and then part of the Langmuir waves convert into electromagnetic waves at the fundamental and harmonic of the local plasma frequency through a non-linear process \citep{reid2014review}. The background electron density decreases with the heliocentric distance, so that the drift from higher frequency to lower frequency observed during a Type-III radio burst corresponds to the fast outward moving electron beams. Using the relationship between the plasma frequency and electron density [${ f_{\rm pe} \rm{[kHz]}=9\sqrt{n_e \rm{[cm^{-3}]}}}$] , one can estimate the heliocentric distance of the wave excitation position with an assumption of the solar and interplanetary background electron density model (\textit{e.g.} \citealp{zhang2018type}).
		
		The positioning using a density model can only yield one dimensional height, which lacks the directional information of the source. The direction of a radio wave train can be obtained by the goniopolarimetric (GP) technique; {which is also referred to as the direction-finding method \citep{bougeret1995waves,cecconi2008stereo}.} {There are two possible ways of implementing GP techniques, one is the spin demodulation based on self-spin stabilized spacecraft such as \textit{Wind} \citep{bougeret1995waves}, \textit{Ulysses} \citep{Stone1992urap} and \textit{Interplanetary Monitoring Platform-6} (IMP-6). The other is the instantaneous GP based on three-axis stabilized spacecraft such as the \textit{Solar Terrestrial Relations Observator} {STEREO} \citep{kaiser2008stereo}.} {GP} has been widely used for studying the source of the Type-III radio-burst. As an example, \cite{fainberg1972radio} deduced the Archimedean spiral nature of the source trajectory using {IMP-6}. The positioning using single spacecraft needs a background electron density model to estimate the heliocentric distance source. However, the source positioning can be achieved directly using the triangulation method with {GP} by multiple spacecraft. \cite{reiner1986new} first applied the {GP} triangulation method to the positioning of Type-III radio burst source. Afterward, the triangulation of the Type-III burst source using \textit{Wind} and \textit{Ulysses} supported that the trajectory of a Type III radio source follows the spiral shape of interplanetary magnetic-field \citep{reiner1998type}. The spiral-shaped magnetic-field line configuration is also referred to as the Parker spiral \citep{parker1958dynamics}. Recently, \cite{krupar2012goniopolarimetric} introduced the single value decomposition (SVD) algorithm into {GP}, so that {GP} can handle extended sources better and estimate the source size. Using this method, \cite{krupar2014statistical} did a statistical survey for the source location of 153 solar Type-III bursts observed by the {STEREO} spacecraft. The result also indicated that the trajectory of a Type-III radio source has a good agreement with the Parker spiral field lines. They found that a source can be largely extended to tens of degrees for the lowest frequency of the events. This indicates there may exist scattering of wave and density fluctuation in interplanetary space.
		
		The {GP} triangulation uses GP-formed data. Moreover, the triangulation for a single point must use the same frequency channel from multiple spacecraft. This means that the {GP} triangulation can only utilize the common frequency band for different spacecraft, while for a radio-burst event, the common band for multiple spacecraft is usually limited to a few frequency channels. 
		
		In this work, we propose a forward-modeling method to estimate the velocity and trajectory of the solar Type-III radio burst source. The method uses the arrival times of the radio waves at multiple spacecraft, which can be obtained from the dynamic spectra of the Type-III burst. The dynamic spectrum form of data is provided by most of the interplanetary spacecraft such as {STEREO} and \textit{Wind}. The \textit{Wind} spacecraft stays near the first Lagrangian point of the Earth since November 1996. The {STEREO} twin spacecraft were launched to provide multiple perspectives of the Sun \citep{kaiser2008stereo}. These two spacecraft operate near Earth's orbit, {STEREO-A (STA)} is inside the Earth's orbit and {STEREO-B (STB)} is outside the Earth's orbit. {STA} and {STB} separate from each other at a rate of about 44$^\circ$ per year. For the time when {STA, STB}, and \textit{Wind} have a large separation angle, the difference of arrival times of radio waves at the three spacecraft can be several minutes depending on the radio source position. The temporal resolution of \textit{Wind}/WAVES and {STEREO/WAVES} is sufficient to distinguish such time differences.
		
		As will be seen below, the proposed forward modeling method makes use of the whole dynamic spectrum of a Type-III burst observed by different spacecraft, which may be in different frequency channels and bands. To achieve the forward modeling, an evaluation function is essential, which measures how well the model fits the observation. The ``best fit model'' makes the modeled object closest to the observation. For a Type-III radio-burst, assuming the radio source is moving along the Parker (Archimedes) spiral, we can model the frequency drift pattern observed by each spacecraft for a given velocity and injection position of the exciter-electron beam. The forward modeling process obtains a parameter set that can make the modeled frequency-drift curves close enough to the leading edge of the observed pattern. The ``best fit'' parameter set finding process in this work is handled by a computational method named the particle swarm optimization (PSO) algorithm \citep{eberhart1995new,shi2001particle}.
		
		This article is arranged as follows: in Section 2, we present the assumptions of the model. In Section 3, the details of the data processing are introduced. In Section 4, we choose five IP Type-III radio-burst events, which either are located by {GP} triangulation or have well-defined origin. We display the forward-modeling results of these events, comparing the estimated results with previous studies. Section 5 is the conclusion and discussion.
		
		\section{Model} 
		
\begin{figure}  
			
			\centering
			\includegraphics[width=12cm]{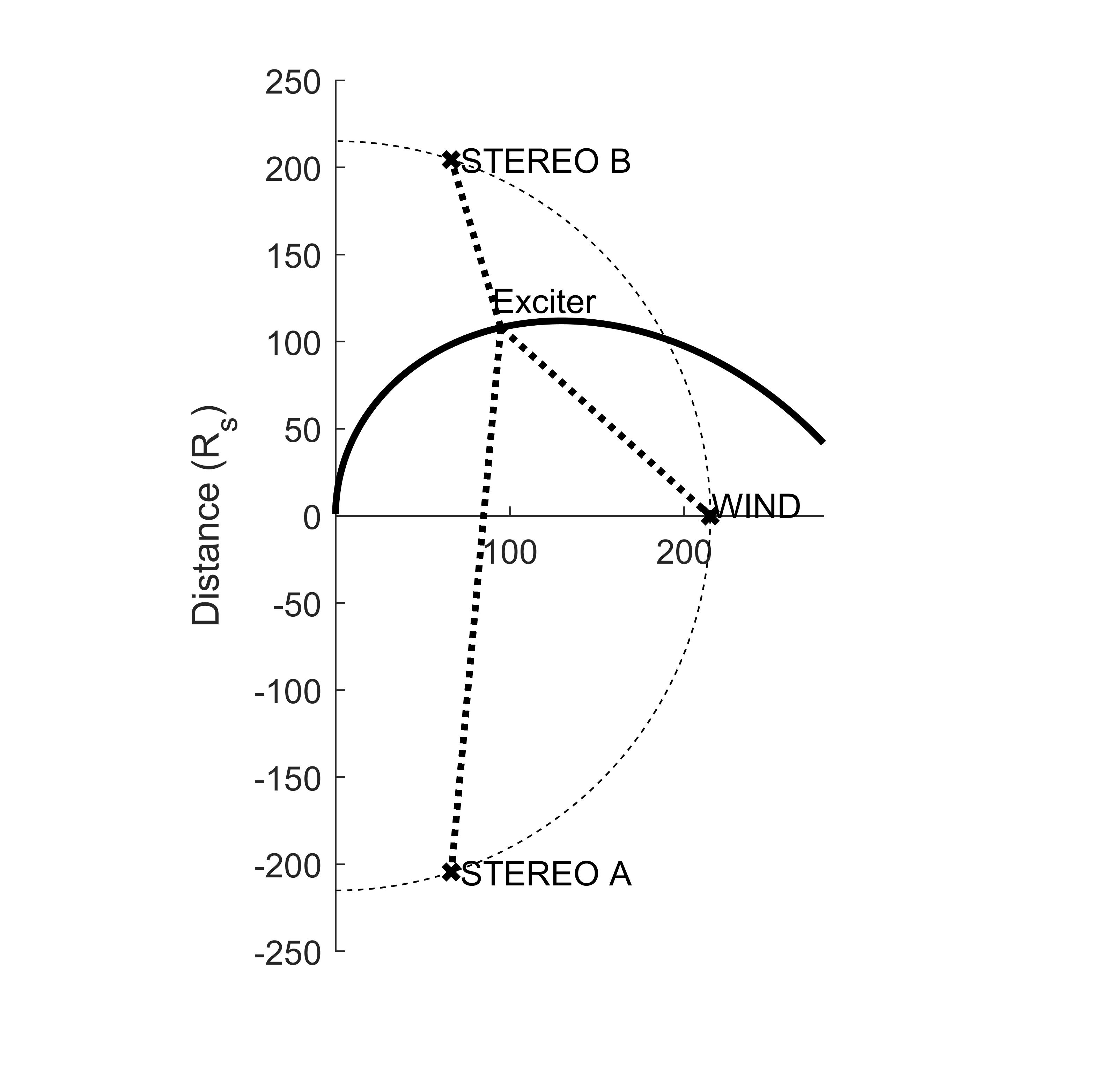}
			\caption{A sketch illustrating the trajectory of the radio source in the Ecliptic plane (\textit{solid line}) and propagation paths of the radio waves to the spacecraft, {STEREO} A, B, and \textit{Wind} (\textit{dotted lines}).}
			\label{fig:1}
		\end{figure}

		The purpose of this work is to forward model the trajectory and velocity of an IP Type-III radio-burst emission source using the arrival time of the radio waves at multiple spacecraft. The arrival time [${t_{\rm a}}$] of the waves at frequency [${f}$] includes three parts (Figure \ref{fig:1}), 
		\begin{equation}
		t_{\rm a}(f) = t_0+t_{\rm m}(f)+t_{\rm p}(f),
		\label{eq:1}
		\end{equation}
		where $t_0$ is the injection time of the electron beam near the solar surface, $t_{\rm m}$ is the motion time of the beam from the injection position to the position where the radio waves are excited, namely, the source position of the radio waves at the corresponding frequency [$f$], and ${t_{\rm p}}$ is the propagating time of the radio wave from the wave-excitation position to the spacecraft receiver. In Figure \ref{fig:1}, a polar coordinate system is used in the ecliptic plane with the origin at the center of the Sun. The source of the IP Type-III radio burst is assumed to move outward from the Sun following the Parker spiral field line \citep{dulk1987speeds,reiner2015electron},

		\begin{equation}
		r=r_0-b(\theta-\theta_0),
		\label{eq:2}
		\end{equation}
		where $r$ is the heliocentric distance from the Sun, $\theta$ is the longitudinal angle displacement from the Sun--Earth line. The angle $\theta_0$ is the longitude of the injection position of the electron beam near the solar surface at $r_0$. The coefficient $b$ is defined as $b=v_{\rm sw}/\Omega$, where $\Omega$ is the angular velocity of the solar rotation and $v_{\rm{sw}}$ is the solar wind speed. From Equation \ref{eq:2}, we can derive the motion distance (arc length) of the electron beam by integrating along the Parker spiral \citep{reiner2015electron},
		\begin{equation}
		S(r)=\frac{r}{2} \sqrt{1+\left(\frac{r}{b}\right)^2 }+\frac{b}{2}  \ln{ \left( \frac{r}{b}+\sqrt{1+\left(\frac{r}{b}\right)^2} \right) }  .
		\end{equation}
		The motion time [$t_{\rm m}$] can be expressed as
		\begin{equation}
		t_{\rm m} = \frac{S(r)-S(r_0)}{v_{\rm s}} ,
		\end{equation}
		where $v_{\rm s}$ is the {average} speed of the radio source along the Parker spiral. {The exciter may experience deceleration in IP space \citep{krupar2015speed,reiner2015electron}, which is ignored in this work for simplicity.} We also assume that there is no significant refraction or scattering of the radio wave during its propagation from the source position to the spacecraft receiver. In other words, the wave propagates along the straight line from the source to the receiver at the speed of light. The wave propagation time [$t_{\rm p}$] in Equation \ref{eq:1} is
		\begin{equation}
		t_{\rm p}=\frac{d_i (r)}{\rm c}= \frac{1}{\rm c} \sqrt{(r \cos\theta-R_i \cos \alpha_i )^2+ (r \sin\theta-R_i  \sin\alpha_i  )^2 } ,
		\end{equation}
		where $d_i(r)$ is the distance between the radio source and the $i$th spacecraft, c is the speed of light. The coordinate points $(r,\theta)$ and $(R_i,\alpha_i)$ are the heliocentric distance and longitude of the radio source and the $i$th spacecraft, respectively. 
		
		According to the plasma emission mechanism, the beam electrons generate radio waves at the fundamental or harmonic of the local plasma frequency. Given an interplanetary electron density model [$n_e (r)$], we can convert the radial distance to a frequency. The arrival time of the radio wave at frequency $f$  can be expressed as
		\begin{equation}
		t_{\rm a}(f) = t_0+\frac{S(f)-S(r_0)}{v_{\rm s}}+\frac{d_i(f)}{\rm c} ,
		\label{eq:tf}
		\end{equation}
		where $f=f_{\rm pe}(r)$ for fundamental emission and $f=2f_{\rm pe}(r)$ for the second harmonic. In this work, we use the density model obtained by \cite{leblanc1998tracing} multiplied by a constant coefficient:
		\begin{equation}
		n_{\rm e} = c_{n} (2.8\times 10^5 r^{-2}+3.5\times 10^6  r^{-4}+6.8\times 10^7 r^{-6}) ,
		\label{eq:ne}
		\end{equation}
		where $r$ is in units of solar radius, $n_{\rm e}$ is in units of cm$^{-3}$, and $c_{\rm n}$ is a constant coefficient. This density model will be referred as Leblanc98 in the following. The time--frequency curve described by Equation \ref{eq:tf} is the modeled leading edge of the dynamic spectrum observed by the $i$th spacecraft. The starting time and frequency drift rate of the time--frequency curve is determined by the electron injection time [$t_0$], the velocity of the exciter or radio source [$v_{\rm s}$], the longitude of the event on the solar surface [$\theta_0$], and the solar wind speed [$v_{\rm sw}$]. We need to point out that, using Equation \ref{eq:tf} for at least three spacecraft can yield the source position from common frequency channel of these spacecraft, theoretically. However, there are few common frequency channels for multiple spacecraft, and the frequency band for a Type-III event received by different spacecraft is usually different. So we introduce the density model to make use of the complete frequency-drift pattern received by all of the spacecraft.
		
\begin{figure}  
			\centering
			\includegraphics[width=12cm]{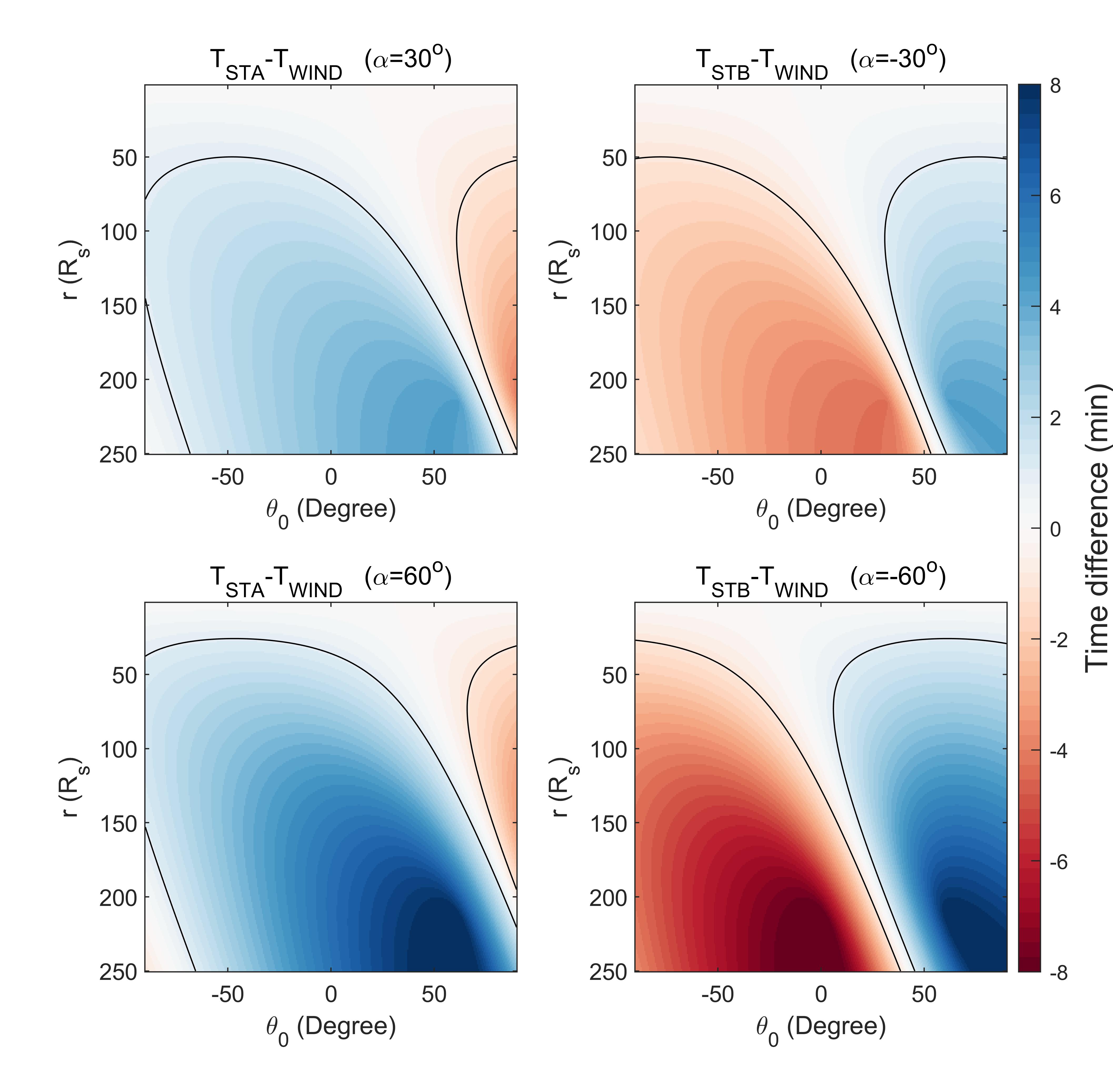}
			\caption{Contour plots of the difference between the radio signal arrival time at {STEREO} and \textit{Wind} for the same Type-III burst, \textit{versus} the injection longitude [$\theta_0$] of the electron beam and the heliocentric distance of the radio source. \textbf{The upper and lower panels} represent the results when the separation angle between the Earth and the {STEREO} are $\pm$ 30$^\circ$ and $\pm$ 60$^\circ$, respectively. \textbf{The left and the right panels} represent the time difference from \textit{Wind} to the STA and STB, respectively. The \textit{black line} in each panel shows the contour line where the time difference is equal to the temporal resolution of spectra used in this work (one minute). The solar wind speed used in the calculation is 400\,km s$^{-1}$.
			}
			\label{fig:2}
		\end{figure}

		Using Equation \ref{eq:tf}, we can calculate the difference of the arrival time of the radio wave at different spacecraft as shown in Figure \ref{fig:2}. The result indicates that the larger the separation angle between the {STEREO} spacecraft and \textit{Wind} is, the larger the arrival time difference can be. Thus, the large separation angle between the spacecraft is conductive to the positioning of the source. The difference of the arrival time generally increases with the heliocentric distance of the radio source. When the radio source is far away from the Sun, namely more than tens of solar radii, the time difference can be large enough to be distinguished in the observational data, depending upon the spacecraft separation angle. Moreover, one can see that when the heliocentric distance of the radio source is larger than tens of solar radii, the indistinguishable regions in left and right panels do not overlap. This means that arrival time difference can be distinguished by at least two of the three spacecraft. In brief summary, the time difference of the signal arriving at different spacecraft can be distinguished in the temporal resolution of observed dynamic spectra with proper locations of the spacecraft.
		
		To achieve the forward modeling, we use the penalty function as the evaluation function, which is defined by,
		\begin{equation}
		D_{\rm total}(t_0,\theta_0,v_{\rm s})= \sqrt{\frac{1}{N}\sum_i \sum_j w_{ij}
			\left(T_i(f_j)-t_i(f_j)\right)^2} ,
		\label{eq:8}
		\end{equation}
		where $i$ and $j$ are the index of the spacecraft and the frequency channel, respectively, $T_i$ is the observed arrival time of the signal at the spacecraft, $t_i$ is the modeled arrival time according to Equation \ref{eq:tf}, $N$ is the total number of selected arrival time entries, and $w_{ij}$ is the weight of each channel. The weight factor $w_{ij}$ indicates the quality of the data. In present work, the dynamic spectra of both \textit{Wind}/WAVES and {STEREO/WAVES} have a temporal resolution of one minute and hundreds of frequency channels in the range of about 10\,kHz\,--\,15\,MHz, so that the same weight is chosen for all data points. However, the weight factor would be useful if observation data with different temporal-frequency resolutions and signal-to-noise level is introduced. We can assign larger weight to the data with higher quality.
		
\begin{figure}   
			\centering
			\includegraphics[width=12cm]{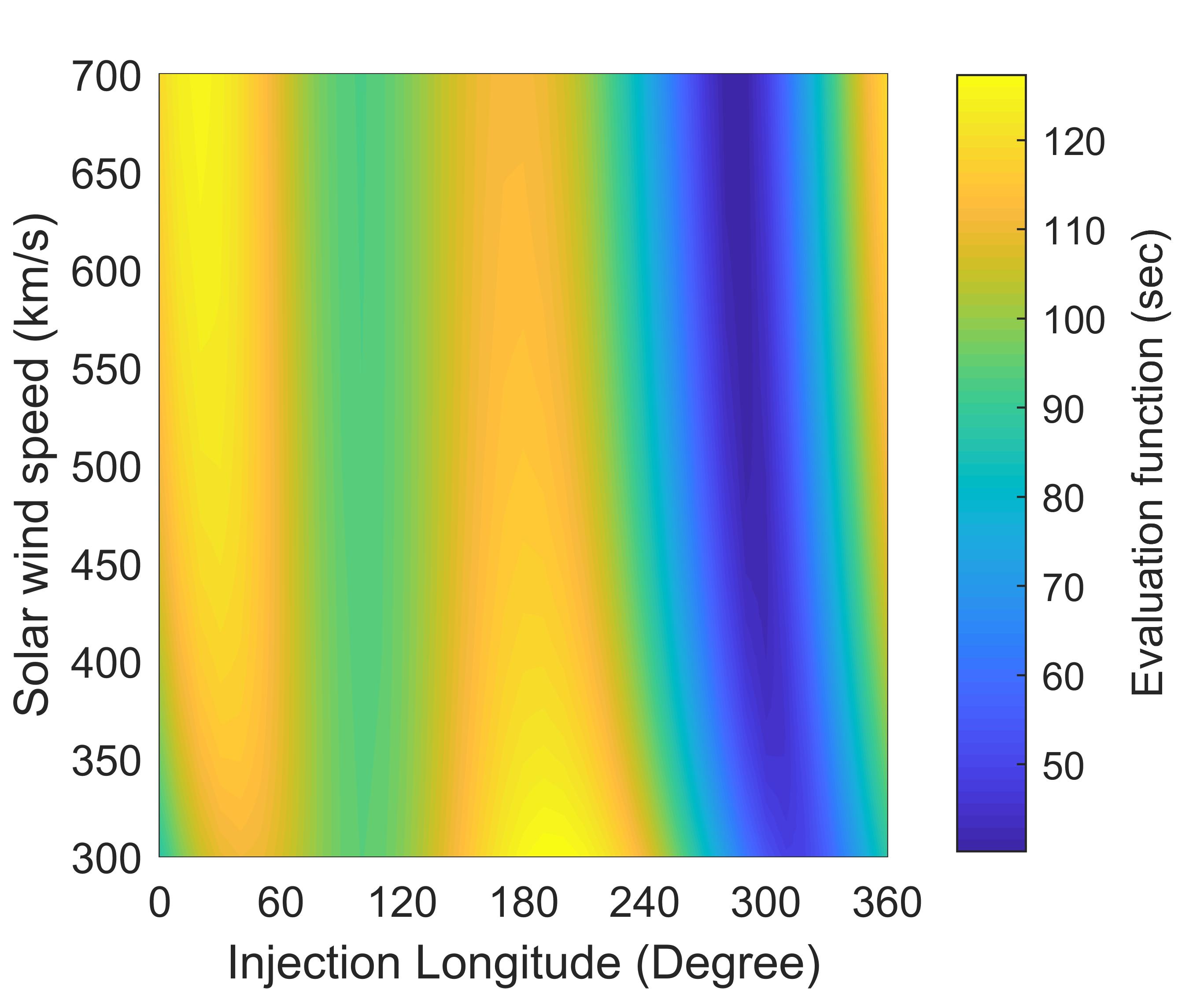}
			\caption{The value of evaluation function with different injection-longitude angles and solar-wind speeds for the case on 29 January 2008.
			}
			\label{fig:9}
		\end{figure}

		The evaluation function describes the disparity between the model and observation. The target parameters of the model are $t_0,\theta_0, v_{\rm  s}$, which represents the injection time of the electron beams, the longitude of the event at the solar surface, and the speed of the radio source along the Parker spiral. {Here, the solar-wind speed is fixed with a constant value of 400\,m s$^{-1}$, since we find that the optimized results are not sensitive to the solar wind speed (see Figure \ref{fig:9} and the following discussion in Section 5).} The forward-modeling process is to alter these parameters in the expression $t_i(f_j)$ of Equation \ref{eq:tf} to minimize the evaluation function. The most direct and precise way to find an extreme point of a function is by differentiation, which can yield an analytical expression of the extreme point. When the function has several variables and the expression is complex, it is more appropriate to use numerical methods. In this work, the PSO algorithm is used to determine the ``best fit model''. The PSO algorithm optimizes the evaluation process by improving the evaluation candidate with regard to the evaluated points \citep{shi2001particle}. It can avoid local extreme-points and dramatically reduce the global extreme point searching time in an extreme-point searching problem. Moreover, new channels and new spacecraft data can be simply introduced by appending the arrival time and the spacecraft position to the evaluation function.
		
		\section{Data Processing} 
		In this work, we use the dynamic spectra observed by {STEREO/WAVES} and \textit{Wind}/WAVES for the same IP Type-III radio-burst. The {STEREO/WAVES} has three radio receivers, the \textit{Low Frequency Receiver} (LFR), the \textit{High Frequency Receiver }(HFR), and the \textit{Fixed Frequency Receiver}(FFR) at 30\,MHz . The {LFR} and {HFR} receive signals in the frequency ranges of 2.5\,--\,160\,kHz and 125\,--\,16025\,kHz, or in other words, the signals of radio waves from the Sun in the heliocentric distance region of about 0.5\,--\,1\,AU and 0.01\,--\,0.5\,AU, respectively. The Wind spacecraft locates near the first Lagrangian point since November 1996. The radio receiver \textit{Wind}/WAVES yields dynamic spectra in two bands. The radio receiver band one (RAD1) and two (RAD2) covers the frequency range of 20\,--\,1040\,kHz and 1.075\,--\,13.825\,MHz, respectively. Both RAD1 and RAD2 have 256 frequency channels. In the present work, we use the dynamic spectra of the {LFR} and {HFR} from {STEREO/WAVES} and RAD1, RAD2 from  \textit{Wind}/WAVES. The temporal resolution of the dynamic spectra we use is one minute. The data are downloaded from NASA's data website \url{cdaweb.gsfc.nasa.gov} in the form of Common Data Format (CDF) files. Only the relative intensity is used to locate the leading edge of the spectrum.
		
\begin{figure}    
			\centering
			\includegraphics[width=12cm]{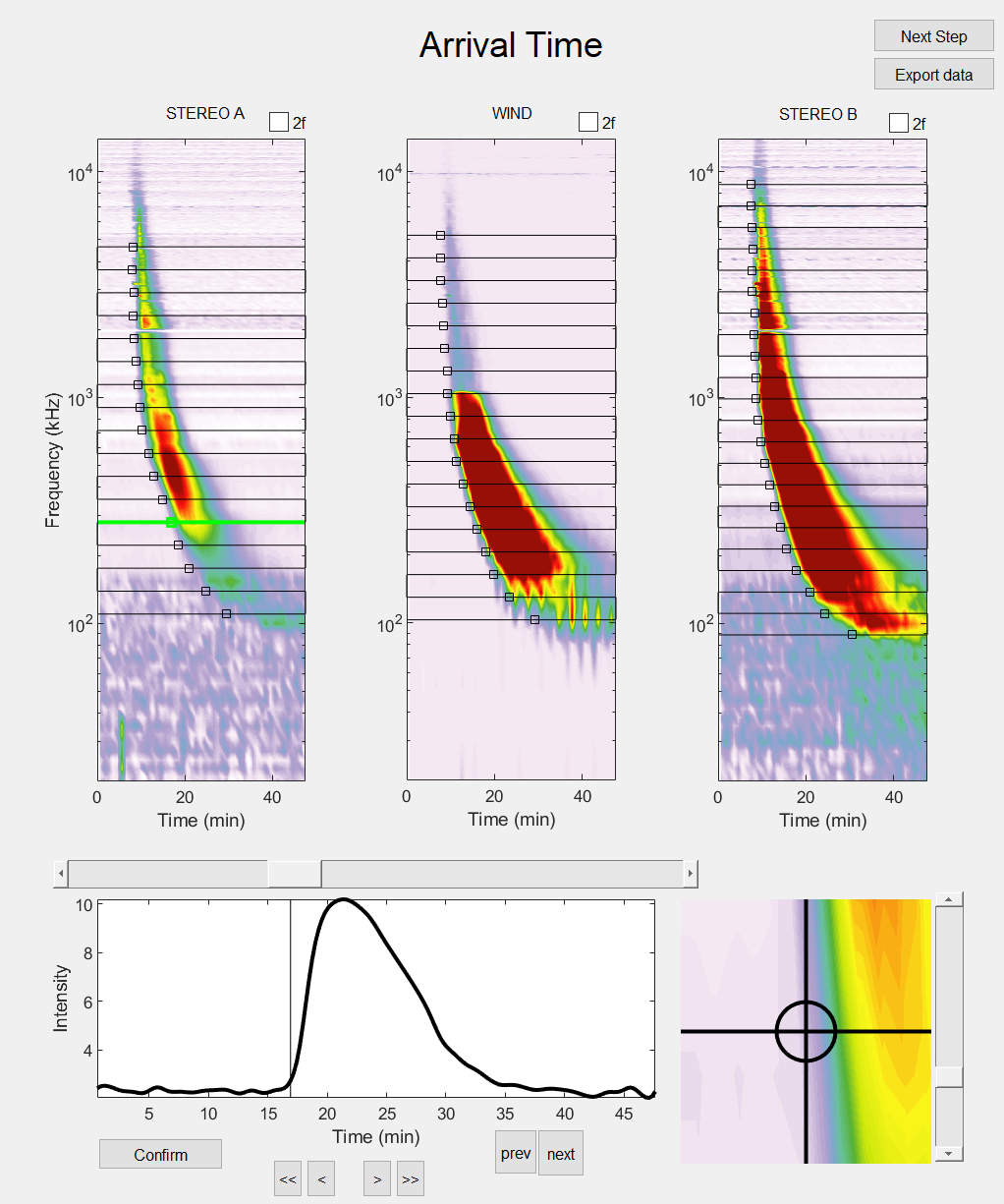}
			\caption{The user interface layout of the program for manually marking of the arrival time.\textbf{ The upper three panels} show the dynamic spectra of the event on 29 January 2008 observed by three spacecraft, where the \textit{black horizontal lines} represent the channels selected for marking. Usually, the frequency channels with clear signal are selected for the marking of the arrival time. The \textit{green line} denotes the current marking channel. \textbf{The lower-left panel} shows the light curve of the current channel. One can use the scrollbar to mark the time of the arrival of the radio burst, and make some slight adjustment using the left and right keys on the keyboard. \textbf{The lower-right panel} is a zoom-in of the dynamic spectra near the selected point. This zoom-in is provided for user to make sure the arrival time is determined accurately.
			}
			\label{fig:3}
		\end{figure}

		The forward-modeling program is composed of three components: First, an arrival-time acquirer is designed to determine the arrival time of radio waves at different frequency channels from the dynamic spectra observed by different spacecraft. A user-friendly graphic user interface (GUI) [{{ Source code updated at [\url{github.com/Pjer-zhang/SEMP}] }}] is designed to import and read the CDF format data file, select the temporal ranges and frequency channels of interest, then manually mark the arrival time of the signal. After selecting the frequency channels where the leading edge of the event is sharp and clear, the light curve of the selected frequency channel is interpolated from the dynamic spectrum. The main layout of the GUI for marking the arrival time on the light curve is shown in Figure \ref{fig:3}. One can mark the arrival time of the selected channel by moving the slider to the beginning time of the burst on the light curve, as shown on the lower-left panel in Figure \ref{fig:3}. For this event, one can see that the light curve contains the burst in the temporal range from 17 to 34 minutes, and there are weak fluctuations before 17 minutes. We use the earliest time when the intensity in the light curve is larger than the maximum value of the fluctuating intensities before the event as the arrival time of the burst signal. The user can make some slight adjustments according to the zoom-in plot in the lower-right panel in the GUI using the left and right keys, to make sure that the marked arrival time is aligned with the leading edge of the Type-III burst dynamic spectra.

		The second component is the optimizer, which minimizes the evaluation function defined by Equation \ref{eq:8} with the PSO algorithm. The input of the optimizer is the arrival time of the selected frequency channels from the three spacecraft. The output is the optimized parameter set including $t_0,\theta_0,v_{\rm s}$, namely, the injection time of the electron-exciter beam, the longitude of the event at the solar surface, and the speed of the radio source (or exciter) moving along the Parker spiral.	
		
		The third component is the error analyzer, which examines the deviation between the model and the observation. Let
		\begin{equation}
		dt_i (f_j) = T_i (f_j)-t_i (f_j)
		\end{equation}
		which represents the difference between the modeled arrival time and the observed one at the $i$th spacecraft for the $j$th frequency channel. To check the performance of the model, we also calculate the maximum of the difference between the modeled arrival time at different spacecraft, which is defined by
		\begin{equation}
		Dt_{\rm max} (f_j )=\max\left[\left| t_{i_1} (f_j )-t_{i_2}  (f_j ) \right| \right]   ,
		\end{equation}
		where $t_{i_1}$ and $t_{i_2 }$ are the modeled arrival time at the spacecraft $i_1$ and  $i_2$, respectively. The value of  $Dt_{\rm max}$ can help us evaluate the possibility that the event could be distinguished by the three spacecraft. The larger the value of $Dt_{\rm max}$ comparing with the instrument temporal resolution, the more possible the burst could be distinguished. Moreover, the relative value of $dt_i$ and $Dt_{\rm max}$ can offer a measurement on the reliability of modeled results. The smaller the value of $dt_i$ compared with that of $Dt_{\rm max}$, the more reliable the corresponding forward-modeling results would be. If the time difference between the model and observation [$dt_i$] is larger than the maximum of the modeled time difference [$Dt_{\rm max}$] for most of the selected frequency channels, the modeled result may not be reliable.
		
		\section{Case Analysis}
		To test the performance of the forward-modeling method described above, we apply it to four IP Type-III solar radio bursts that have been well discussed in the literature. {The date and the frequency range of the four events are shown in Table \ref{T1}. The first three columns show the frequency range of the STA, \textit{Wind} and {STB}, the fourth column shows the common frequency range. The selected frequencies are marked in the dynamic spectra figure for each event.} These events are selected to benchmark the forward modeling method.
		
		\begin{table}
			\caption{ The date and frequency range of the four events. The first column  of each data cell is the observed frequency range, the second column is the frequency range used for forward modeling. The selected frequency range is in the unit of kHz}
			\label{T1}
			\begin{tabular}{lcccc}
				\hline
				Date & STA              & WIND             & STB              & Common Frequency \\
				\hline
				\multirow{2}{*}{2008.01.29} & 80kHz$\sim$10MHz & 100kHz$\sim$8MHz & 30kHz$\sim$12MHz & 100kHz$\sim$8MHz \\
				& {[}110,4646{]}   & {[}103,5216{]}   & {[}89,8768{]}    & {[}110,4646{]}   \\
				\multirow{2}{*}{2010.01.17} & 30kHz$\sim$1MHz  & 50kHz$\sim$3MHz  & 90kHz$\sim$12MHz & 90kHz$\sim$1MHz  \\
				& {[}54,1068{]}    & {[}101,966{]}    & {[}89,8201{]}    & {[}101,966{]}    \\
				\multirow{2}{*}{2010.11.17} & 60kHz$\sim$12MHz & 30kHz$\sim$5MHz  & 200kHz$\sim$2MHz & 200kHz$\sim$2MHz \\
				& {[}87,3140{]}    & {[}69,1835{]}    & {[}225,1627{]}   & {[}225,1627{]}   \\
				\multirow{2}{*}{2011.11.03} & 20kHz$\sim$4MHz  & 70kHz$\sim$1MHz  & 40kHz$\sim$10MHz & 70kHz$\sim$1MHz  \\
				& {[}42,3060{]}    & {[}83,874{]}     & {[}50,6277{]}    & {[}83,874{]}    \\
				\hline
			\end{tabular}
		\end{table}

		\subsection{Event on 29 January 2008}
\begin{figure}    
			\centering
			\includegraphics[width=12cm]{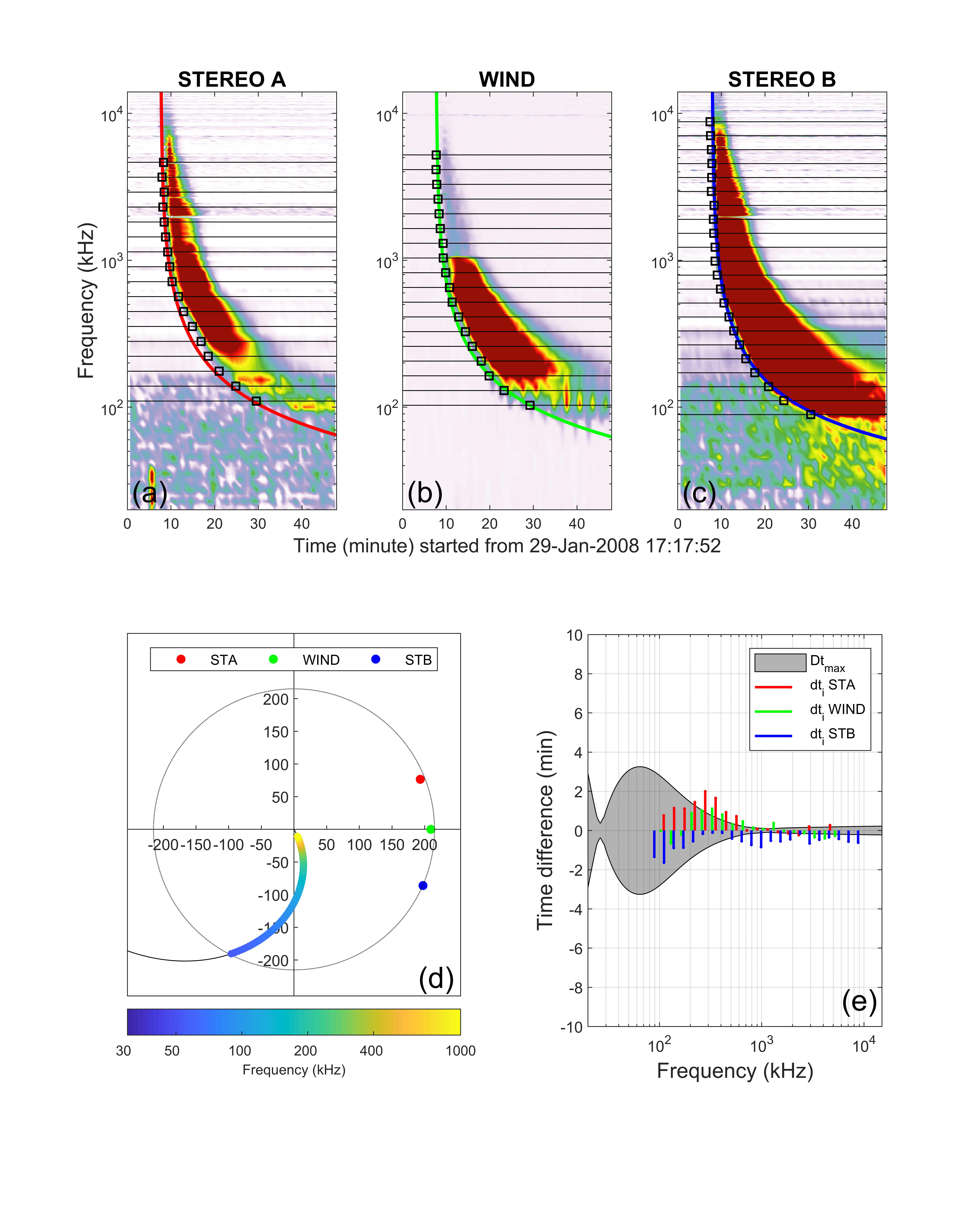}
			\caption{The event on 29 January 2008. \textbf{Panels a, b, c} are the dynamic spectra from {STEREO} and \textit{Wind}, \textit{the horizontal black line} denotes the selected frequency channels. The \textit{squares} show the manually marked arrival time. \textit{The colored lines} show the modeled frequency drift curve or the leading edges. \textbf{Panel d} shows the modeled trajectory of the source, where the positions of the source are colored according to the local plasma frequency. \textbf{Panel e} shows the deviation between the model and observation, where the \textit{gray area} represents $Dt_{\rm max} (f_{j})$.\textit{ The bars colored with red, green and blue} represent $dt_{i} (f_{j})$ for {STA, \textit{Wind}，} and {STB} respectively.
			}
			\label{fig:4}
		\end{figure}

		This event is observed by {STEREO} and \textit{Wind}/WAVES at 17:25\,UT on 29 January 2008. At the time of the event, STB was located at 1.0015\,AU from the Sun and 23.5\,$^\circ$ behind the Earth. {STA} was at 0.9667\,AU from the Sun and 21.7\,$^\circ$ ahead of the Earth. The {\textit{Wind}}--Sun distance was {0.975}\,AU. The Type-III event covers the frequency range from about 100\,kHz to 10\,MHz. The event is single and clear on the dynamic spectra observed by the three spacecraft, as shown in Figure \ref{fig:4} a\,--\,c. In this case, the waves received by {STA, STB}, and \textit{Wind} are all assumed to be fundamental waves. For the background electron density model, we have tested several values for the coefficient $c_{n}$ and eventually find that six times the Leblanc98 density model can minimize the evaluation function. The forward-modeling result on the parameter $t_0, \theta_0,$ and $ v_{\rm s}$ shows that the estimated speed of the burst source along the Parker spiral is 0.22 c, the longitude of the burst event is 60.5$^\circ$ {east of the \textit{Wind}\,--\,Sun line at the solar surface, and the injection time of the energetic electrons is at 17:17:18\,UT}. With these estimated parameters [$t_0, \theta_0, v_{\rm s}$], we can obtain a modeled time\,--\,frequency curve for each spacecraft. The solid lines in Figure \ref{fig:4} a\,--\,c denote the modeled leading edge with fundamental wave assumption for STA, \textit{Wind} and STB. From Figure \ref{fig:4} a\,--\,c we can see the modeled curve aligns well with the leading-edge of the burst on the dynamic spectra of every spacecraft. Figure \ref{fig:4}d shows the reconstructed trajectory of this burst event with E$60.5^\circ$ injection longitude and 400\,km s$^{-1}$ solar wind speed. The colors on the trajectory denote the local plasma frequency derived from the background electron density model.

		The gray area in Figure \ref{fig:4}e represents the modeled maximum time difference of the arrival time at different spacecraft ($\pm Dt_{\rm max}$). The red, green, and blue bars represent the time deviation between the modeled frequency drift curve and the observed leading edge, which is expressed in Equation \ref{eq:tf}, for {STA}, \textit{Wind} and {STB}, respectively. The separation angle between {STA} and \textit{Wind} or {STB} and \textit{Wind} is about 22$^\circ$. For this event, the maximum value of  $Dt_{\rm max}$ is about three minutes near the frequency 60\,kHz as shown in Figure \ref{fig:4}e. However,  $Dt_{\rm max}$ is smaller than one minute or the instrument temporal resolution for wave frequency greater than 300\,kHz. This is because the source positions of radio wave with these high frequencies are located very close to the Sun. The difference of the wave arrival time to three spacecraft that are all operating near 1\,AU is relatively small for waves from the small heliocentric distance regime, and \textit{vice versa}. From Figure \ref{fig:4}e, one can see that $dt_i$ is within the gray area for the frequency band less than 200\,kHz. {Although the $Dt_{\rm max}$ is smaller than one minute for high frequency channels, we chose several frequency channels above 1\,MHz. The reason is that the leading edge of high frequency is clearer and the arrival time is easier to mark. Moreover, the arrival time of high frequency channels can help us determine the event starting time [$t_0$], which contributes to the final result of the forward modeling.} The evaluation function describes the mean time deviation between the observed leading edge and the modeled frequency drift line. In this case, the final iteration value of the evaluation function is 43 seconds, which is below the temporal resolution of the dynamic spectra observed by {STEREO/WAVES} and \textit{Wind}/WAVES.
		
		This event has been studied by \cite{reiner2009multipoint} {and \cite{martinez2012determination}} using the {GP} triangulation method. This event is considered to be associated with a B1.2 X-ray flare located at 59$^\circ$ east of the Earth. Our estimated injection longitude is about 1.5$^\circ$ east to the center of the associated flare, which is well consistent with the longitude of the active region. \cite{reiner2009multipoint} used the three spacecraft to triangulate the radio source of this event at the frequency channel of 425\,kHz. It is found that the source of signal at 425\,kHz is located 64$^\circ$ east of the \textit{Wind}--Sun line and 0.19\,AU from the Sun. {\cite{martinez2012determination} used {STA} and {STB} to triangulate the source of this event, the source of 425\,kHz is located 73\,$^\circ$ east of the Earth and 0.21\,AU from the Sun.} From our result, the source of 425\,kHz locates 65$^\circ$ east of the \textit{Wind}--Sun line, which is consistent with the result of {GP} triangulation. The heliocentric distance of the wave with plasma frequency of 425\,kHz is about 0.13\,AU according to the electron density model used in this case. This displacement may be produced by the inaccuracy of the density model or by the scattering of the wave due to density inhomogeneities. The leading edge of the dynamic spectrum is used in the forward modeling, while the triangulation mainly investigates the position of the waves with peak intensity on the light curve. The wave of the leading edge arrives at the spacecraft at the earliest time, which is usually less influenced by wave-scattering effects, compared with the waves with the same frequency but arriving later.
		
		The source position of this burst at the frequency channels between 125\,kHz and 1975\,kHz have been triangulated by \cite{krupar2014statistical} using {STEREO}. They found that, for most of the frequency channels, the locations of the sources are on the west side seen from the Earth near the flare site. The event longitude of our forward modeling result is qualitatively consistent with the result of triangulation.

		\subsection{Event on 17 January 2010}
\begin{figure}    
			\centering
			\includegraphics[width=12cm]{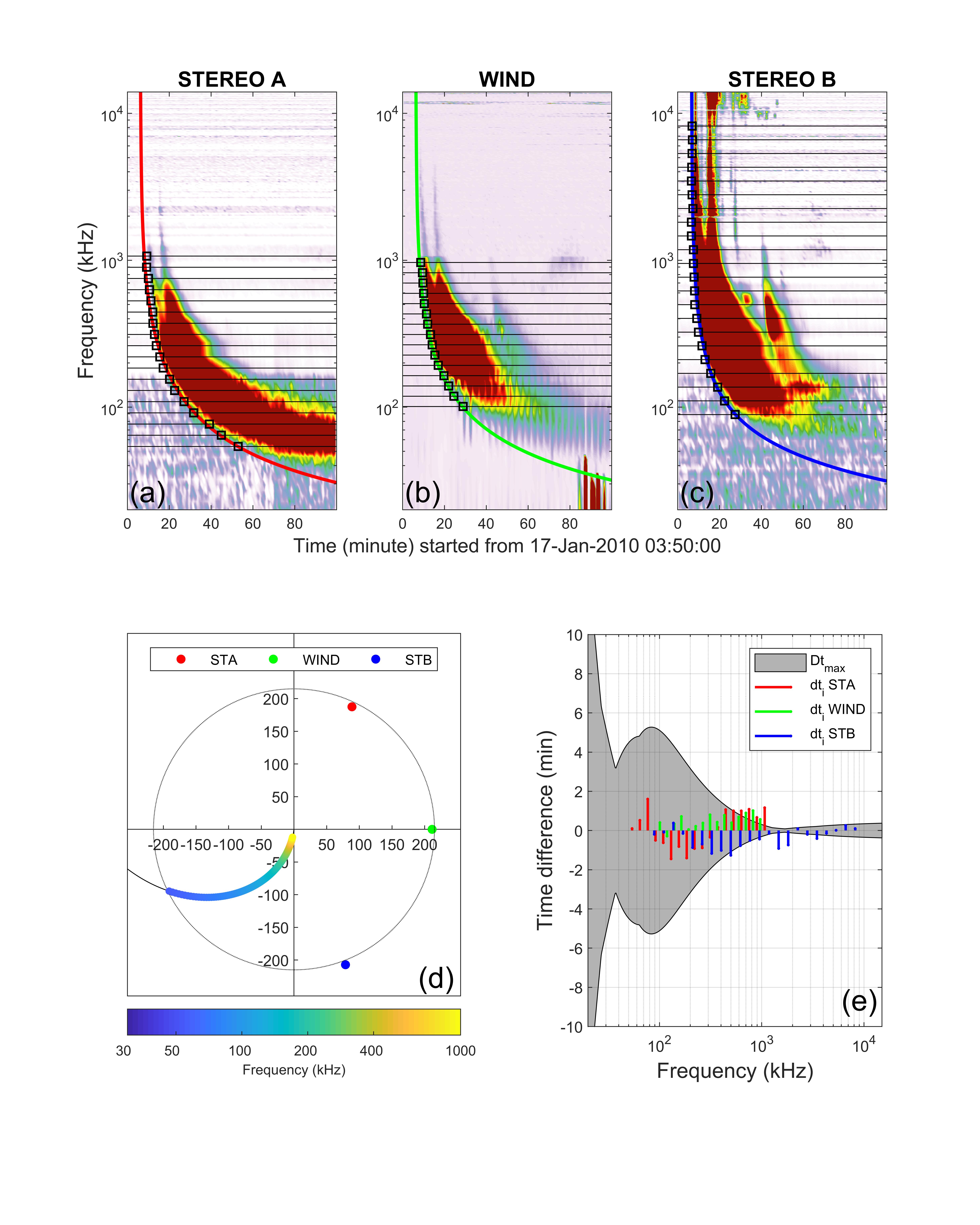}
			\caption{
				The event on 17 January 2010. \textbf{Panels a, b, c} are the dynamic spectra from {STEREO} and \textit{Wind}, \textit{the horizontal black line} denotes the selected frequency channels. The \textit{squares} show the manually marked arrival time. \textit{The colored lines} show the modeled frequency drift curve or the leading edges. \textbf{Panel d} shows the modeled trajectory of the source, where the positions of the source are colored according to the local plasma frequency. \textbf{Panel e} shows the deviation between the model and observation, where the \textit{gray area} represents $Dt_{\rm max} (f_{j})$.\textit{ The bars colored with red, green and blue} represent $dt_{i} (f_{j})$ for {STA, \textit{Wind}，} and {STB} respectively.
			}
			\label{fig:5}
		\end{figure}

		At the time of this event, STB was at 69.2$^\circ$ east of the Earth, and {STA} was at 64.7$^\circ$ west of the Earth. {STB, STA} and \textit{Wind} are located 1.0309\,AU, 0.9649\,AU and {0.973}\,AU from the Sun, respectively. From the dynamic spectra shown in Figure \ref{fig:5}a\,--\,c, we can see that the leading edges of the dynamic spectra are clear. The arrival time can be marked easily. The dynamic spectra from the three spacecraft show there are more than one Type-III burst within the time interval. We assume that in the selected time inverval, the earliest leading edges obtained by the three spacecraft are all from the same source, which is the source of earliest Type-III burst among these Type-III bursts. The leading edge observed by {STA, STB,} and \textit{Wind} are all assumed to be the fundamental wave. We tried to change the coefficient $c_n$  of Equation \ref{eq:ne} from 0.5 to 10, and found that the evaluation function decreases continuously when $c_n$ increases from 0.5 to 5, but it is stable when $c_n$ varies between 5 and 10. We simply choose six times the Leblanc98 density model for this event, the same as that for the event on 29 January 2008. From the result of the forward modeling, the longitude of this event at the solar surface is 97.4$^\circ$ east of the \textit{Wind}--Sun line. The modeled trajectory is shown in Figure \ref{fig:5}d. The speed of the source is 0.24c along the Parker spiral field line, and {the injection time of the electron beam is at 03:47:49\,UT.}
		
		In this case, the separation angles between the event longitude and spacecraft are larger than the event on 29 January 2008. As a result, the difference of arrival time at different spacecraft is larger than that of the above event. It can be seen from Figure \ref{fig:5}e that for most of the frequency band,  $Dt_{\rm max}$ is larger than the value for the event on 29 January 2008. The maximum value of $Dt_{\rm max}$  reaches five minutes near the wave frequency of 90\,kHz. The absolute value of $Dt_{\rm max}$ is larger than two minutes for the selected frequency channels in the frequency range 50\,kHz\,--\,300\,kHz. The time difference between the observed arrival time and the modeled leading edge [$dt_i$] is less than two minutes, which is acceptable considering the temporal resolution of the observation. Thus, in this case the value of $dt_i$ is significantly smaller compared to $Dt_{\rm max}$, as shown in Figure \ref{fig:5}e. The minimum value of the evaluation function is 44 seconds, which is below the temporal resolution of the dynamic spectra observed by STEREO/WAVES and \textit{Wind}/WAVES.
		
		This event has been discussed by \cite{dresing2012large}. There were two active regions (AR) on the solar surface at the time of this Type-III radio burst, namely AR1 (NOAA AR 11039) and AR2 (NOAA AR 11040). The center of AR1 was located at about 127$^\circ$ east relative to the Earth, and AR2 was located at about 60$^\circ$ west relative to the Earth. During the event, AR2 was in its quiet stage, while there was a flare in AR1 at 3:49\,UT. Combined with coronagraph observation, \cite{dresing2012large} inferred that AR1 may be the source of the Type III radio-burst. For comparison, our estimated longitude of the burst source near the solar surface was about 30$^\circ$ west of the center of AR1 and 157$^\circ$ east of AR2. Thus, our result prefers AR1 as the source of this burst, which is consistent with the inference of \cite{dresing2012large}.
		
		\subsection{Event on 17 November 2010}
		
\begin{figure}    
			\centering
			\includegraphics[width=12cm]{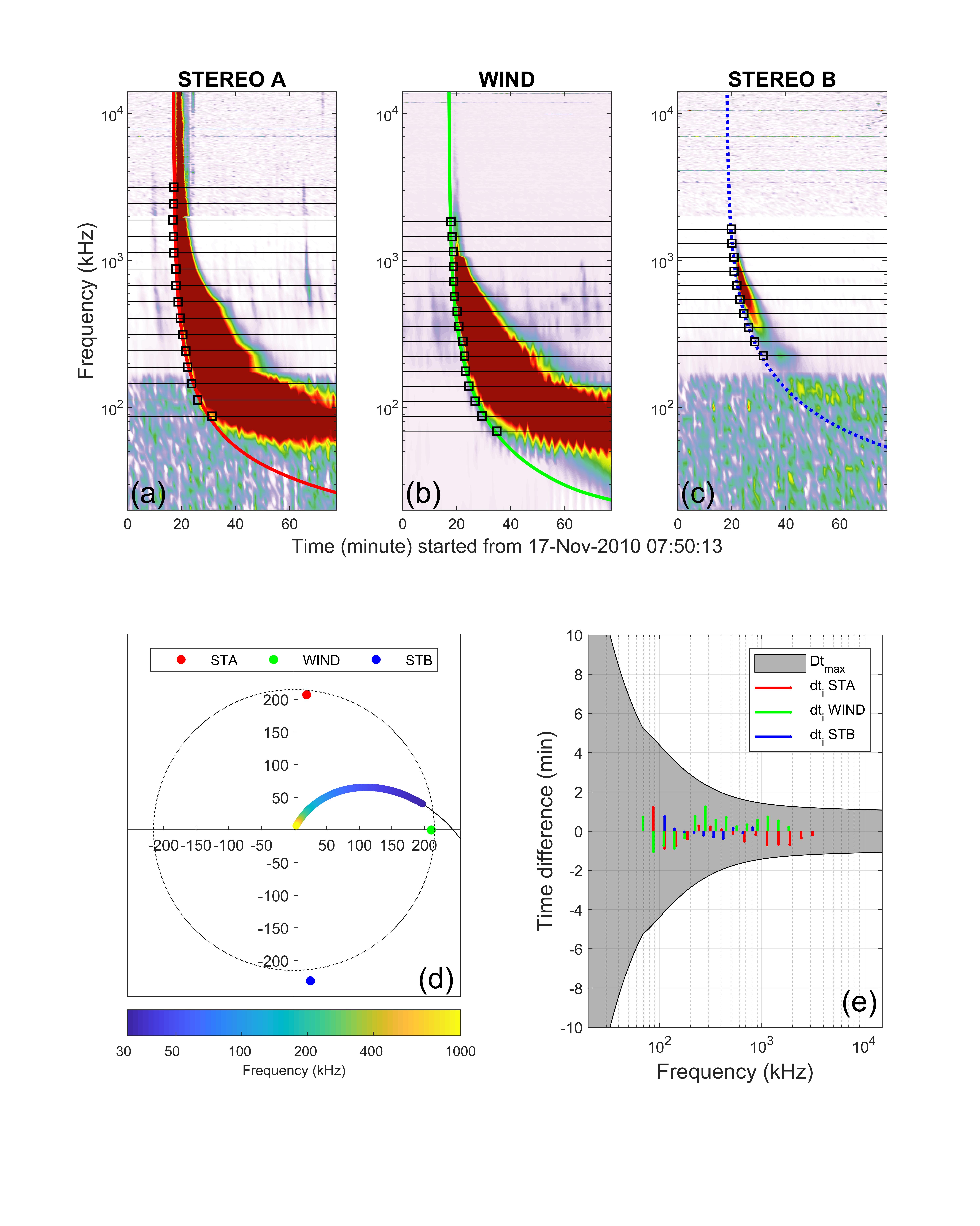}
			\caption{
				The event on 17 November 2010. \textbf{Panels a, b, c} are the dynamic spectra from {STEREO} and \textit{Wind}, \textit{the horizontal black line} denotes the selected frequency channels. The \textit{squares} show the manually marked arrival time. \textit{The colored lines} show the modeled frequency drift curve with the assumption of fundamental wave (\textit{solid lines}) and second harmonic wave (\textit{dashed lines}). \textbf{Panel d} shows the modeled trajectory of the source, where the positions of the source are colored according to the local plasma frequency. \textbf{Panel e} shows the deviation between the model and observation, where the \textit{gray area} represents $Dt_{\rm max} (f_{j})$.\textit{ The bars colored with red, green and blue} represent $dt_{i} (f_{j})$ for STA,\textit{ Wind，} and {STB} respectively.
			}
			\label{fig:10}
		\end{figure}

		{At the time of this event, {STB} was located at 1.081\,AU from the Sun and 83.80$^\circ$ behind the Earth. {STA} was at 0.967\,AU from the Sun and 84.63$^\circ$ ahead of the from Earth. \textit{Wind} was 0.978\,AU from the Sun. This event is strong in the dynamic spectra compared to the background noise. As shown in Figure \ref{fig:10}a\,--\,c, the leading edge of this event is clear and easy to mark. For the forward modeling, we assume the leading edge received by {STA} and \textit{Wind} is the fundamental wave, and that received by {STB}  is the second harmonic. We tried different values of $c_n$ of Equation \ref{eq:ne} from 0.5 to 10, and finally used 2.3 time the Leblanc98 density model according to the evaluation function. The result of the forward modeling shows that the longitude of this event is 64.6$^\circ$ west of the \textit{Wind}--Sun line, the injection of the electron beam happens at 07:58:54\,UT at the Sun, the average speed of the electron beam is about 0.21\,c.}

		{On 17 November 2010, the separation angle in this case is large enough to distinguish the arrival time difference of the signal for multiple spacecraft. The $Dt_{\rm max}$ of this case is shown in Figure \ref{fig:10}e. For the frequency band of the selected frequency channels from 60\,kHz to 2\,MHz, the $Dt_{\rm max}$ is larger than the temporal resolution of the dynamic spectra of {STEREO} and \textit{Wind}. For the low frequency band from 60\,kHz to 100\,kHz, $Dt_{\rm max}$ is more than four minutes. The modeled frequency drift line fits well to the leading edge in the dynamic spectra, as shown in Figure \ref{fig:10}a\,--\,c. The final value of the evaluation function is 34 seconds. The forward modeling result can be considered reliable for this event.}
		
		{This event was studied by \cite{reiner2015electron}. There are Langmuir waves and local plasma emissions observed by the \textit{Wind} spacecraft \citep{reiner2015electron}. An electron event was detected by \textit{Electron, Proton, and Alpha-particle Monitor} (EPAM) instrument on the \textit{Advanced Composition Explorer} (ACE) spacecraft after the Type III burst. The Langmuir waves and the electron event indicate that the electron beam passed near the Earth. From our forward-modeling result, the modeled beam trajectory also passed near the Earth (shown in Figure \ref{fig:10}d). The velocity of the electron beam obtained by analyzing the time of local emission is 0.28\,c \citep{reiner2015electron}, which is slightly faster than the speed given by forward modeling (0.21\,c). The electron beam injection time is 07:58:54\,UT (7.98\,UT) according to the forward modeling, this injection time is consistent with the analysis by \cite{reiner2015electron}, which gives 7.99\,UT as the absolute time of commencement at the Sun.}

		\subsection{Event on 03 November 2011}
\begin{figure}    
			\centering
			\includegraphics[width=12cm]{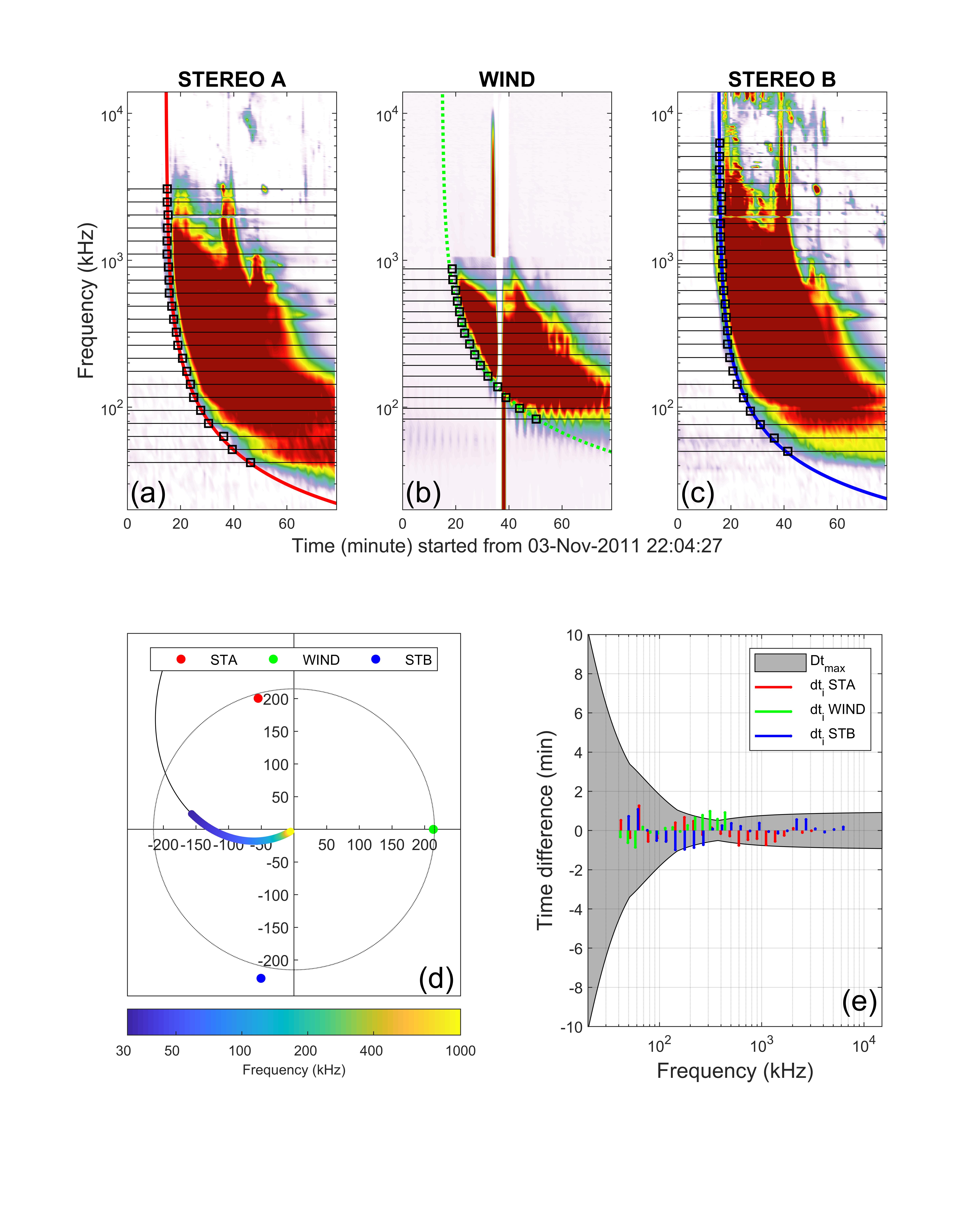}
			\caption{
				The event on 3 November 2011. \textbf{Panels a, b, c} are the dynamic spectra from {STEREO} and \textit{Wind}, \textit{the horizontal black line} denotes the selected frequency channels. The \textit{squares} show the manually marked arrival time. \textit{The colored lines} show the modeled frequency drift curve with the assumption of fundamental wave (\textit{solid lines}) and second harmonic wave (\textit{dashed lines}). \textbf{Panel d} shows the modeled trajectory of the source, where the positions of the source are colored according to the local plasma frequency. \textbf{Panel e} shows the deviation between the model and observation, where the \textit{gray area} represents $Dt_{\rm max} (f_{j})$.\textit{ The bars colored with red, green and blue} represent $dt_{i} (f_{j})$ for STA,\textit{ Wind，} and {STB} respectively.
			}
			\label{fig:7}
		\end{figure}

		At the time of this event, \textit{STA} was located at 105.3$^\circ$ to the west of Earth, and 0.967\,AU away from the Sun. \textit{STB} was located at 102.5$^\circ$ to the east of Earth, and 1.086\,AU away from the Sun. The distance between the Sun and \textit{Wind} was 0.982\,AU. The Type-III burst on the dynamic spectrum is strong, as shown in Figure \ref{fig:7}a\,--\,c. The radio burst lasts over 30 minutes in most of the channels between 50\,kHz and 1.2\,MHz. This event contains multiple Type-III bursts. And there are some spike structures and a type-II burst after the Type-III event. In this study, we focus on the first Type III radio burst which begins at about 20 minutes in Figure\ref{fig:7} a\,--\,c. The leading edge of the first Type III event is clear and easy to mark. For the forward modeling, we assume the leading edges observed by the three spacecraft came from the same source. The leading edges observed by {STA} and {STB} are assumed to be the fundamental wave, while that observed by \textit{Wind} is assumed to be the second harmonic. The density model used is 1.0 times the Leblanc98 model. From the forward modeling result, the speed of the electron beam is 0.16\,c along the Parker spiral field line. {The injection time of the electrons is at 22:10:31\,UT near the solar surface.} The longitude of the event at the solar surface is 147$^\circ$ to the east of Earth. {With the combined observations of {STEREO} and SOHO, this event is identified to be associated with a large two-ribbon flare at N10E50 from the point of view of {STB}, which corresponds to 152$^\circ$ east from the Earth's point of view \citep{gomez2015circumsolar}. The loop brightening started between 22:11 and 22:16\,UT.} The injection longitude of the energetic electrons is consistent with the combined observations.

		In this case, the separation angles are nearly 120$^\circ$ for each pair of spacecraft. The spacecraft are in their most favorable position for forward modeling Type-III burst. From Figure \ref{fig:7}e, we can see that most of the $dt_i$ are smaller than the value of $Dt_{\rm max}$. The final value of the evaluation function is 31 seconds. We consider that the forward-modeling result can be regarded as reliable for this event.
		
		The \textit{in-situ} observation shows that the electron flux observed by {STA} is stronger than that observed by {STB} \citep{gomez2015circumsolar}. According to the source trajectory inferred from the forward modeling, even though the event longitude is closer to {STB} at the solar surface, the source center is closer to {STA} at 1\,AU.  The energetic electrons can be diffused more easily to {STA} than to {STB}. As a result, {STA} would have observed a stronger energetic electron flux although the flare-site is closer to {STB} in longitude.

		\section{Conclusion and Discussion}
		The position of the radio source yields important information about the exciter super-thermal electrons and can help diagnose the associated active regions near the solar surface. Previous radio burst positioning studies in the interplanetary region have used the {GP techniques} \citep{cecconi2008stereo}, which can derive the source of the radio burst independent of background electron-density models. {GP requires measurement by several antennas and multi-channel receivers to perform spin demodulation or instantaneous GP inversion. Moreover,} GP triangulation can only use the common frequency channels of the multiple spacecraft, which is usually limited to a few frequency channels for the triangulation operation. 
		
		In this work, we proposed a forward-modeling method, which uses the arrival time of a Type-III radio burst wave at different spacecraft to estimate the trajectory of the radio source. This method assumes that the exciter moves along the Parker spiral field line in the interplanetary space. With this method, we can derive the outward speed and the injection time, as well as the longitude of the electron beam near the solar surface, which triggers the Type-III radio burst. To automate the arrival time labeling and estimation, we developed a pipeline. The pipeline allows us to mark the leading edge of the event in a user-friendly GUI. Then, the arrival times of the wave at selected frequency channels for different spacecraft are input into the PSO optimizer to finally derive the best-fitting parameters for the Type III radio burst. The target parameters of the PSO optimization are the source speed, injection time, and the longitude of the event near the solar surface. The forward-modeling method that we proposed in this work uses the dynamic spectra data, which is a general form of radio observation. Moreover, it can make use of the whole spectra of an event from multiple spacecraft. This method can help to determine the active region associated with the Type-III radio burst. With a large enough number of events, this method can also be used to study the directivity of the Type III radio-burst.
		
		An implicit assumption of the forward-modeling method is a point source for the radio wave at a given frequency. This kind of assumption is also used in the triangulation of the Type-III radio source \citep{reiner1998type}. Observations indicate that the size of Type-III bursts increases with decreasing frequency. The observed source size of Type-III radio bursts in interplanetary space has been previously measured by \cite{steinberg1984type} and recently reviewed by \cite{reid2014review}. The observed source size is less than one degree near 10\,MHz band and can extend to tens of degrees in the low-frequency channels of {STEREO} and \textit{Wind} (20\,--\,100\,kHz). The size of the radio source can be influenced by two aspects. One is the spatial distribution of the electron-beam exciter. The other is the expansion of apparent source size due to scattering from density inhomogeneities in the background plasma \citep{steinberg1971coronal, kontar2017imaging}. The less scattered waves arrive at the observer earlier, and \textit{vice versa}. This produces an increase in the apparent area of a source with time. In this work, we use the leading edge in the dynamic spectrum. This means that for all of waves excited at the same frequency, we only consider the radio wave that is excited and arrives at the spacecraft at the earliest time. Although the complete extension area of the energetic electrons may be large, the electrons generating the leading edge of the spectrum may be distributed in a limited, small region. Moreover, the leading edge of the radio burst is less influenced by scattering. Thus, we assume a point source in the forward modeling.
		
		An electron-density model of the corona and interplanetary space is needed to determine the heliocentric distance of the radio source. There are various kinds of density models obtained from different observations, most of which are remote sensing. In this work, we used the Leblanc98 model multiplied by a constant coefficient. The Leblanc98 model combines the radio observation from different observatories and covers the range from a few solar radius to 1\,AU (Leblanc, 1998). We find that, from the result of the forward modeling, the source location longitude  [$\theta_0$] at the solar surface is not sensitive to the coefficient value of $c_{\rm n}$. For example,  $\theta_0$ varies within $\pm$ 2$^\circ$ when $c_{\rm n}$ changes from 5 to 9 for the event on 17 January 2010. However, the speed of the beam is highly influenced by the density model. In practice, we chose the $c_{\rm n}$ by minimizing the evaluation function.
		
		In addition, we find the result is not sensitive to the solar wind speed, so the solar wind speed is set as a constant value in the forward modeling process. Figure \ref{fig:9} shows the evaluation function with different injection longitude and solar wind speed for the event on 29 January 2008. We can see that for a given value of $\theta_0$, the value of evaluation function is almost invariable with respect to the solar wind speed changing between {300} and 700\,km s$^{-1}$. {For the solar wind speed of 300\,km s$^{-1}$, the event longitude is 63.9$^\circ$ to the east of Earth, the injection time is 17:17:16\,UT, and the speed of the electron beam is 0.24\,c. For 700\,km s$^{-1}$, the event longitude is 64.5$^\circ$ to the east of Earth, the injection time is 17:17:00\,UT, and the speed of the electron beam is 0.21\,c.} The results of other events are similar. In practice, the solar wind speed is set as 400\,km s$^{-1}$ in this work for the five events introduced above. However, we still encourage the user to choose the solar wind referring to the real solar wind condition to get a more rigorous result. In the future, solar wind speed prediction models such as the Wang--Sheeley--Arge(WSA) model \citep{arge2000improvement} can be integrated into the model to make the forward modeling more self-consistent.

		\begin{acks}
			We acknowledge use of NASA/GSFC's Space Physics Data Facility's  CDAWeb service, and {STEREO,\textit{Wind}} data.The research was supported by the National Nature Science Foundation of China(41574167 and 41174123) and the Fundamental Research Funds for the Central Universities (WK2080000077). Y. Wang is supported by the grants from NSFC (41574165, 41774178 and 41842037). 
		\end{acks}
		
		\section*{Disclosure of Potential Conflicts of Interest}
		The authors declare that they have no conflicts of interest.
		
		\bibliographystyle{spr-mp-sola}

	\end{article} 
	
\end{document}